\documentclass{emulateapj}
\usepackage{apjfonts}
\submitted{Accepted for publication in ApJ (1st April 2013)}

\usepackage{natbib}
\usepackage{epsfig}

 \newcommand{\be}{\begin{equation}}
 \newcommand{\ee}{\end{equation}}

 \newcommand{\kms}{\mbox{km\ \ensuremath{\rm{s}^{-1}}}}
 
\shortauthors{Cordiner et al.}

\begin{document}

\title{On the ubiquity of molecular anions in the dense interstellar medium}

\author{M. A. Cordiner\altaffilmark{1}$^,$\,\altaffilmark{2}, J. V. Buckle\altaffilmark{3}, E. S. Wirstr{\"o}m\altaffilmark{1}$^,$\,\altaffilmark{4}, A. O. H. Olofsson\altaffilmark{4} and S. B. Charnley\altaffilmark{1}}

\altaffiltext{1}{Astrochemistry Laboratory and The Goddard Center for Astrobiology, NASA Goddard Space Flight Center, Code 691,
8800 Greenbelt Road, Greenbelt, MD 20771, USA.}
\email{martin.cordiner@nasa.gov}
\altaffiltext{2}{Department of Physics, The Catholic University of America, Washington, DC 20064, USA.}
\altaffiltext{3}{Cavendish Astrophysics Group and Kavli Institute for Cosmology,
Institute of Astronomy, University of Cambridge, Madingley Road, 
Cambridge, CB3 0HE, UK.}
\altaffiltext{4}{Department of Earth and Space Sciences, Chalmers University of Technology, Onsala Space Observatory, 439 92 Onsala, Sweden.}

\begin{abstract}

Results are presented from a survey for molecular anions in seven nearby Galactic star-forming cores and molecular clouds.  The hydrocarbon anion C$_6$H$^-$ is detected in all seven target sources, including four sources where no anions have been previously detected: L1172, L1389, L1495B and TMC-1C. The C$_6$H$^-$/C$_6$H column density ratio is $\gtrsim1.0$\% in every source, with a mean value of 3.0\% (and standard deviation 0.92\%). Combined with previous detections, our results show that anions are ubiquitous in dense clouds wherever C$_6$H is present. The C$_6$H$^-$/C$_6$H ratio is found to show a positive correlation with molecular hydrogen number density, and with the apparent age of the cloud.   We also report the first detection of C$_4$H$^-$ in TMC-1 (at 4.8$\sigma$ confidence), and derive an anion-to-neutral ratio C$_4$H$^-$/C$_4$H~$= (1.2\pm0.4)\times10^{-5}$ ($=0.0012\pm0.0004$\%). Such a low value compared with C$_6$H$^-$ highlights the need for a revised radiative electron attachment rate for C$_4$H.  Chemical model calculations show that the observed C$_4$H$^-$ could be produced as a result of reactions of oxygen atoms with C$_5$H$^-$ and C$_6$H$^-$.

\end{abstract}


\section{Introduction}

Negative ions (anions) were first discovered in the interstellar medium (ISM) by \citet{mcc06}, who detected the linear hydrocarbon anion C$_6$H$^-$ in \mbox{TMC-1}. This was followed by C$_6$H$^-$ and C$_4$H$^-$ detections in the protostellar core L1527 by \citet{sak07} and \citet{agu08}, respectively, which hinted of a role for anions in the chemistry of star-formation. \citet{gup09} performed the first dedicated survey for C$_6$H$^-$ in twenty-four molecular sources, but detected the anion in only two star-forming clouds, with anion-to-neutral ratios on the order of a few percent, similar to previously-observed values. Interstellar anion detections were confined to the Taurus Molecular Cloud complex until the recent discoveries of C$_6$H$^-$ in the Lupus, Cepheus and Auriga star-forming regions \citep{sak10,cor11}, which proved that anions are widespread in the local ISM, and not an artifact of any particular physical or chemical conditions in the Taurus region.

The possible importance of anions in interstellar chemistry was first discussed by \citet{dal73}, and prospects for the detection of molecular anions using radio astronomy were examined by \citet{sar80}. \citet{her81} argued that large interstellar molecules (including carbon-chain-bearing species) can undergo rapid radiative electron attachment -- as exemplified by the laboratory experiments of \cite{wood80} -- potentially resulting in significant anion abundances in dense molecular clouds. Based on this idea, models for anion chemistry have been successful in reproducing the observed abundances of C$_6$H$^-$ and C$_8$H$^-$ in TMC-1, IRC+10216, L1527 and L1512  \citep{mil07,rem07,har08,cor08,cor12}. However, there is a major discrepancy between the modeled and observed C$_4$H$^-$ anion-to-neutral ratio \citep[see][]{her08}, and the lack of anions in PDRs \citep{agu08} is at variance with the model predictions of \citet{mil07}. Clearly, our understanding of molecular anion chemistry is incomplete. Nevertheless, as demonstrated by \citet{cor12}, anion measurements have the potential to offer insight into interstellar cloud properties due to their reactivity \citep{eic07}, and consequently, their sensitivity to the abundances of gas-phase electrons and atomic C, H and O, and to physical conditions such as cloud density and molecular depletion. Observations of anions in a variety of interstellar environments will be key to a complete understanding of their chemistry. Given the relatively small number of molecular anion detections in the ISM to date, we set out to address the question of just how widespread anions are in interstellar clouds, and to ascertain the behavior of the anion-to-neutral ratio over a range of cloud types and ages.

In this article, results are presented from our survey for C$_6$H$^-$ and C$_6$H in a sample of seven carbon-chain-rich interstellar clouds, protostars and prestellar cores. Detections of C$_6$H$^-$ in two of the sources (L1251A and L1512) were previously reported by \citet{cor11}, and here we report the final results for all seven sources, as well as the first (tentative) detection of C$_4$H$^-$ in TMC-1.

\section{Target Sources}
\label{sec:targets}

\begin{deluxetable*}{lcclcc}
\centering
\tabletypesize{\footnotesize}
\tablecaption{Targets, coordinates and distances\label{tab:sources}}
\tablewidth{0pt}
\tablehead{
Source&RA&Dec&Cloud Type\tablenotemark{a}&Distance& Ref.\\
&(J2000)&(J2000)&&(pc)
}
\startdata
L1172 SMM        &21:02:22.1&+67:54:48 & Star-forming     &440&1\\
L1251A           &22:30:40.4&+75:13:46 & Star-forming     &300&2\\
L1389 (CB17) SMM1&04:04:36.6&+56:56:00 & Protostellar     &250&3\\
L1495B           &04:15:41.8&+28:47:46 & Quiescent        &140&4\\
L1512            &05:04:07.1&+32:43:09 & Star-forming     &140&4\\
TMC-1C           &04:41:35.6&+26:00:21 & Star-forming     &140&4\\
TMC-1 CP         &04:41:41.9&+25:41:27 & Quiescent        &140&4
\enddata
\tablenotetext{a}{See Section \ref{sec:targets} for description of source classification scheme.}
\tablerefs{(1) \citet{vis02}; (2) \citet{kun93}; (3) \citet{lau10}; (4) \citet{eli78}.}
\end{deluxetable*}

\begin{figure*}
\centering
\includegraphics[width=0.9\columnwidth,angle=270]{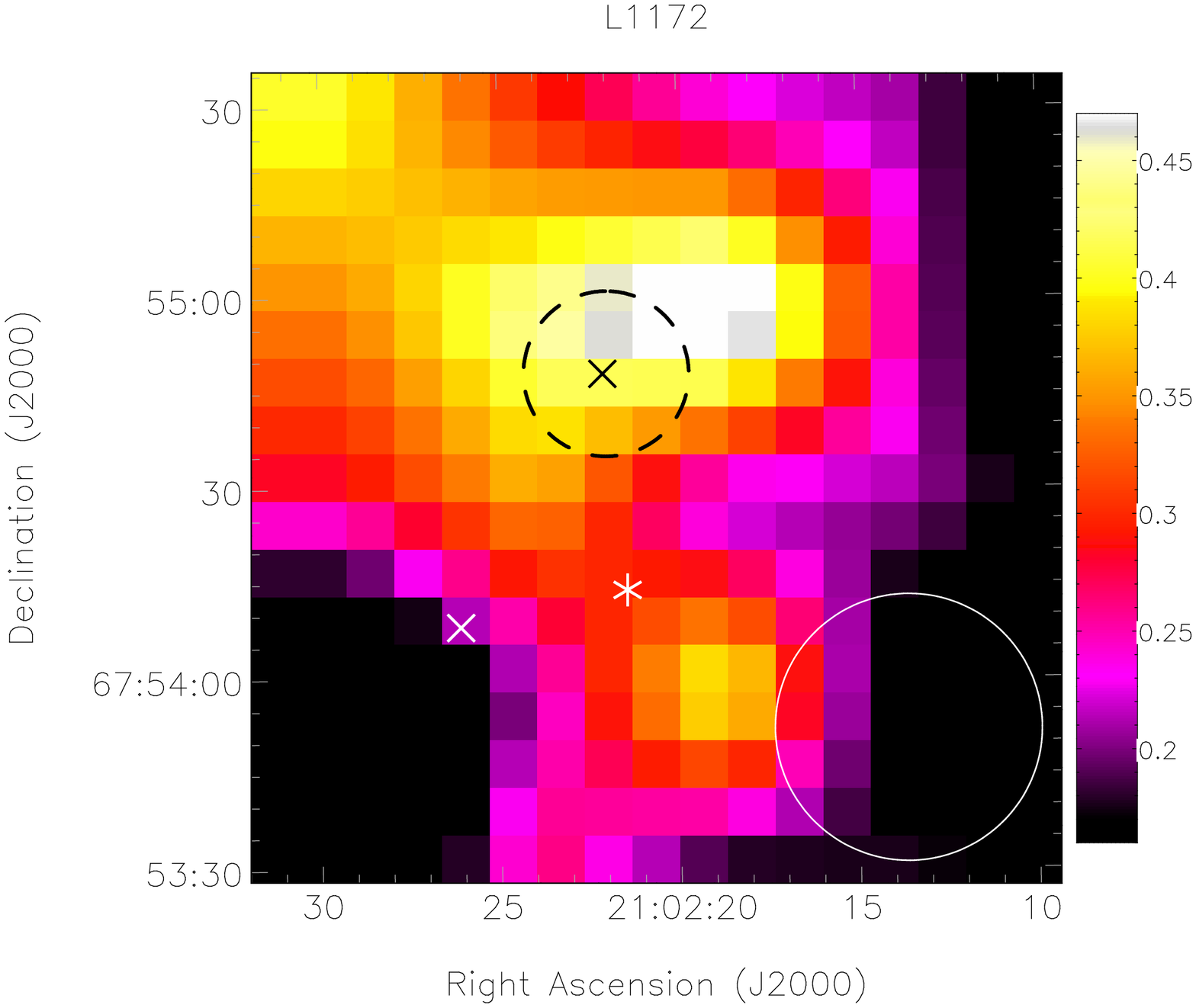}
\includegraphics[width=0.9\columnwidth,angle=270]{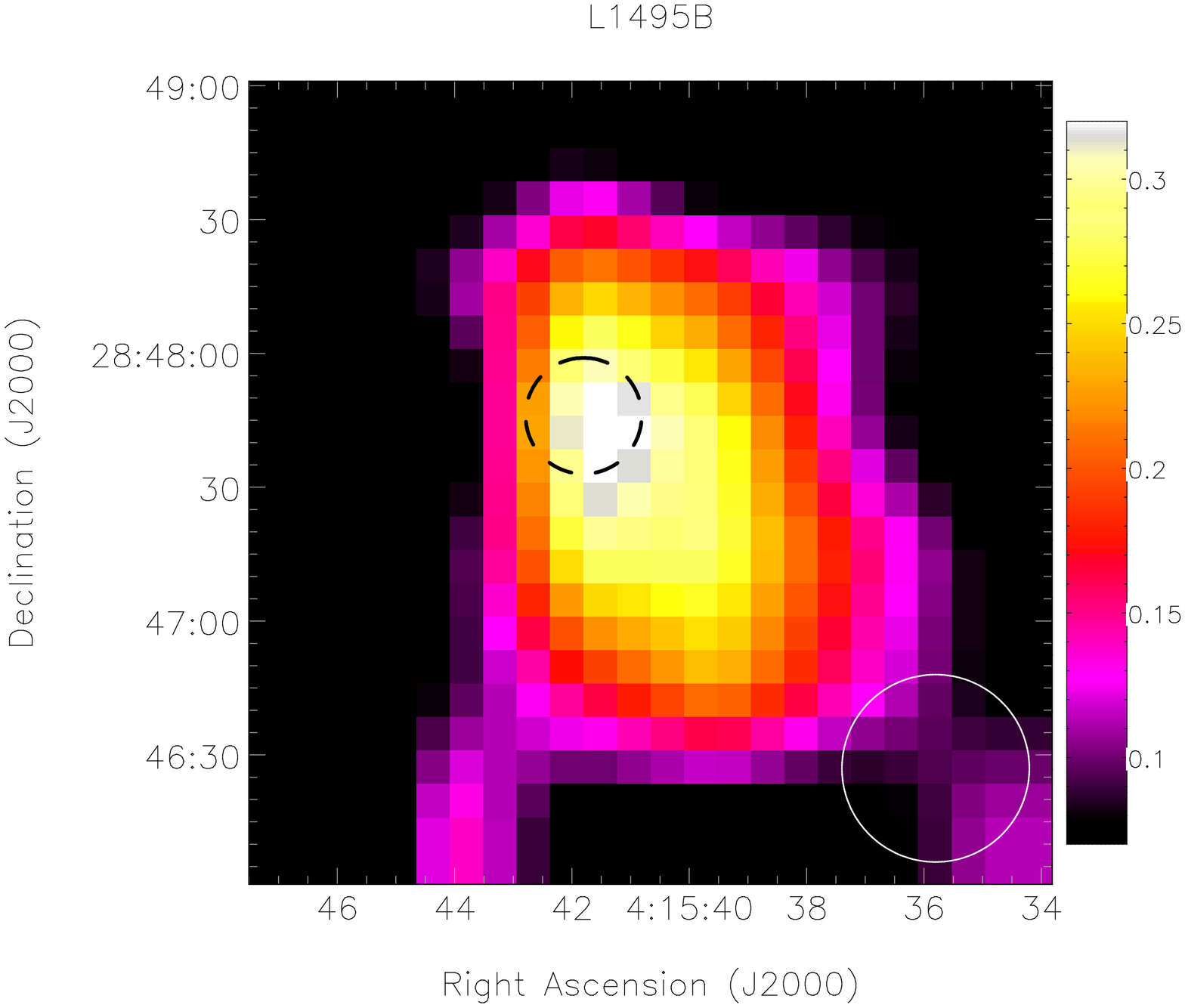}
\includegraphics[width=0.9\columnwidth,angle=270]{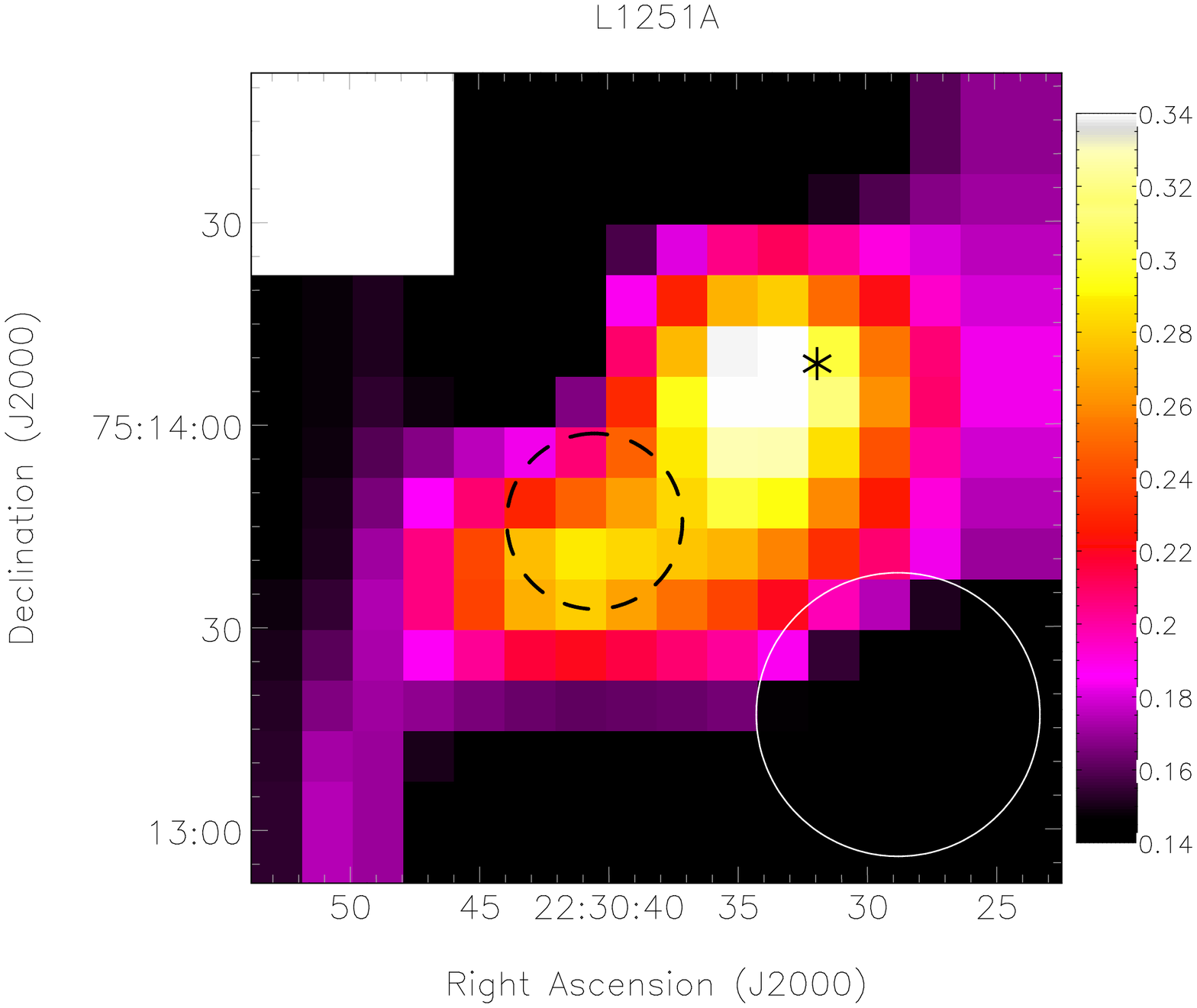}
\includegraphics[width=0.9\columnwidth,angle=270]{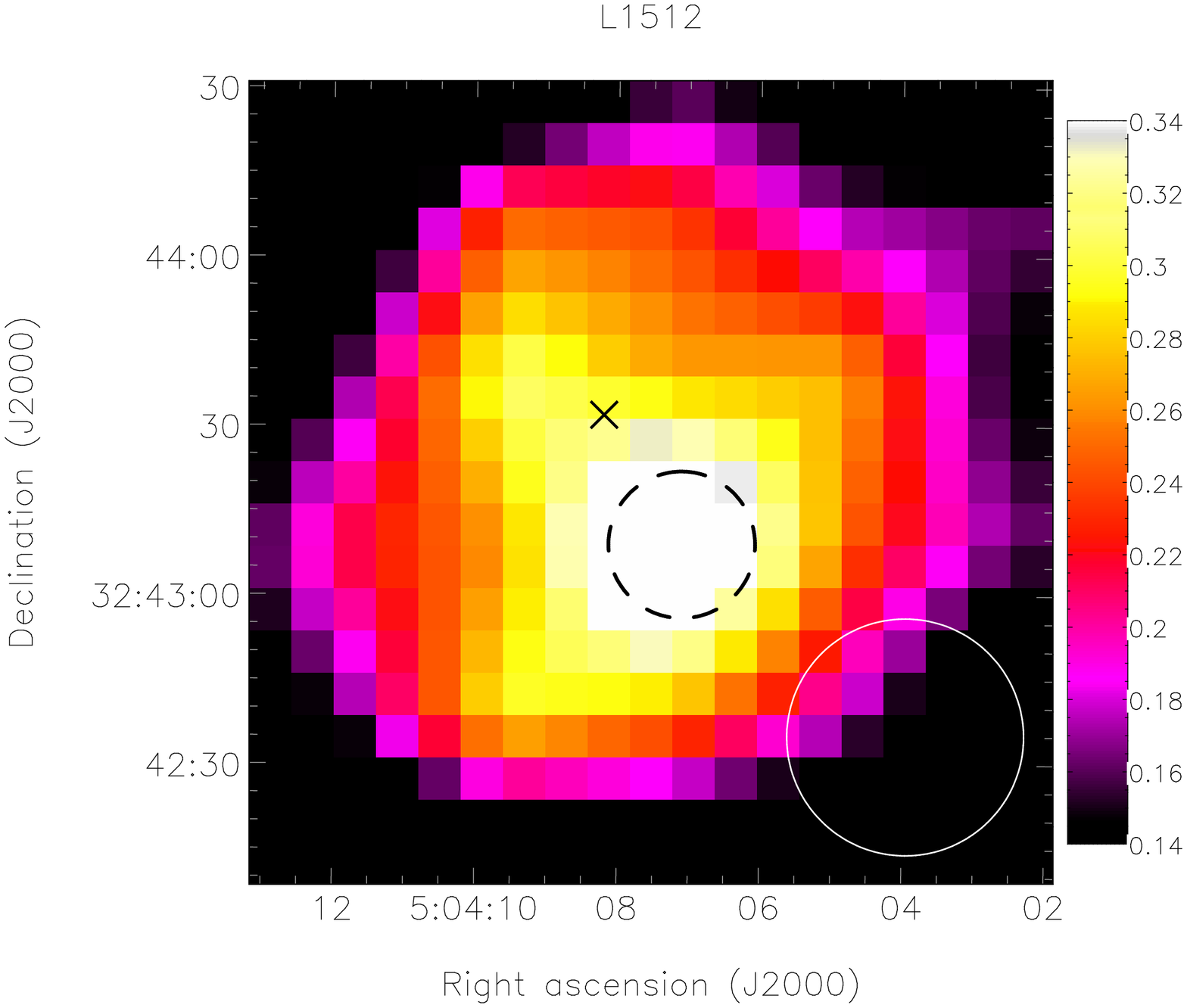}
\includegraphics[width=0.9\columnwidth,angle=270]{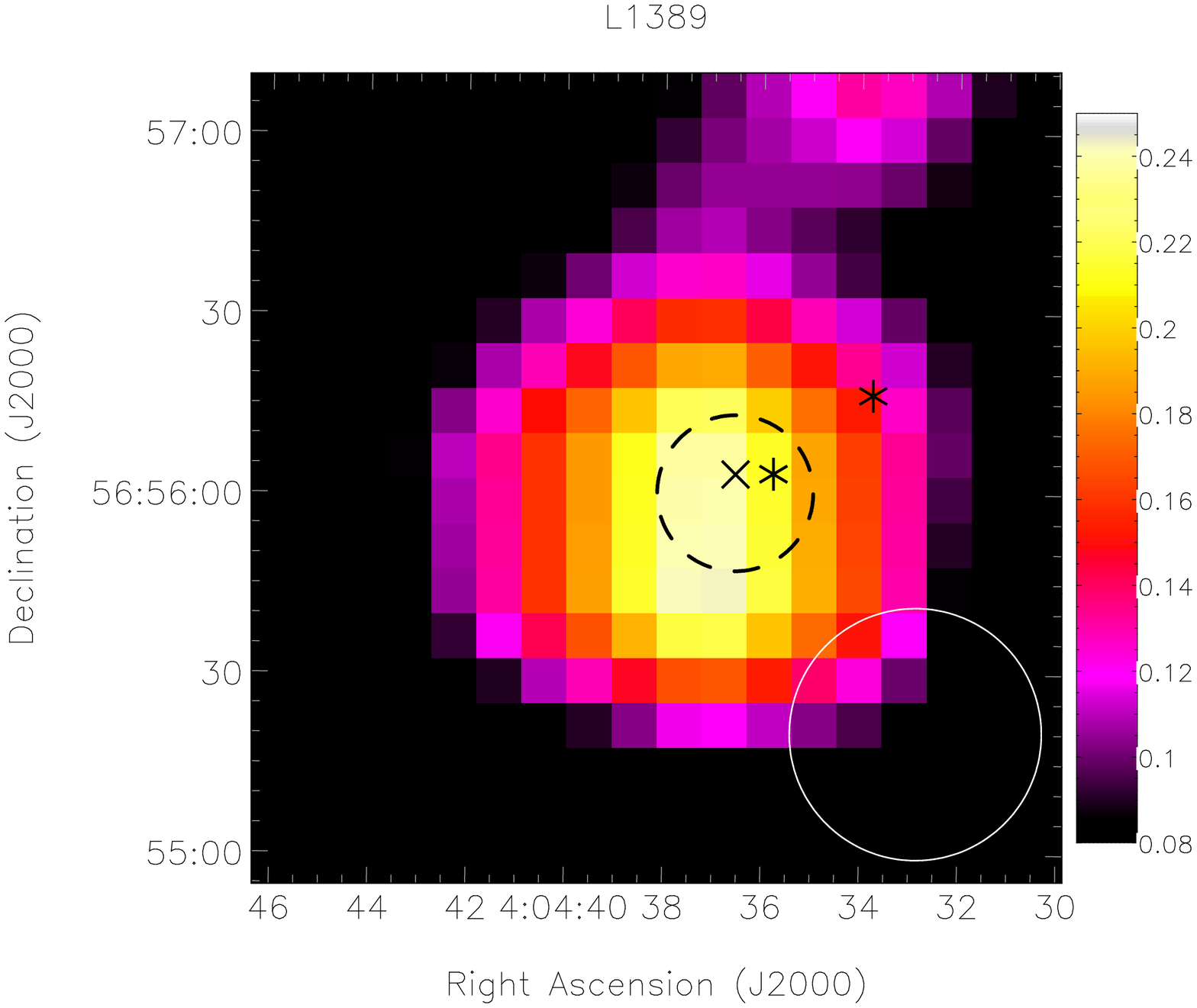}
\includegraphics[width=0.9\columnwidth,angle=270]{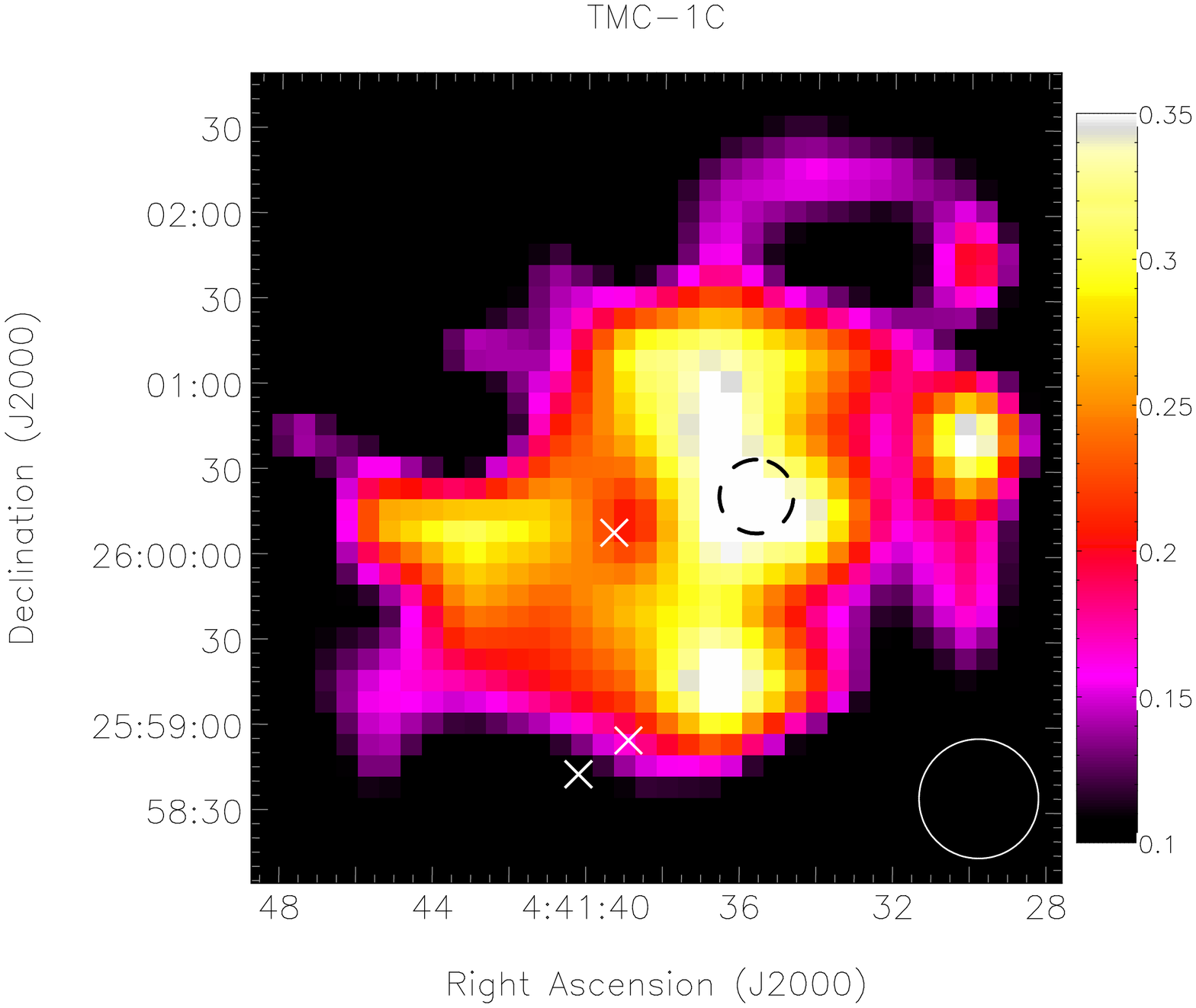}
\caption{Observed HC$_3$N $J=10-9$ integrated intensity maps towards L1172, L1251A and L1389, L1495B, L1512 and TMC-1C. The OSO beam size is indicated in the lower right of each panel with a white circle. Intensity scales are given in units of K\,km\,s$^{-1}$ and have not been corrected for beam efficiency. Dashed black circles represent the size (HPBW) and position of our targeted GBT anion survey positions. Crosses denote locations of sub-mm (prestellar) cores \citep{war94,vis02,dif08,lau10} and asterisks denote protostellar core locations \citep{vis02,che12,lee10}. \label{fig:maps}}
\end{figure*}

Due to the close chemical relationship between polyynes and cyanopolyynes (see \emph{e.g.} \citealt{fed90,mil94}), carbon-chain-bearing species such as C$_4$H and C$_6$H are expected to be abundant in dense molecular clouds, close to the peak of HC$_3$N emission. Target sources with strong HC$_3$N emission lines were selected from an HC$_3$N $J=10-9$ mapping survey of twenty nearby molecular clouds, prestellar cores and young, low-mass protostars obtained using the Onsala Space Observatory (OSO) 20-m telescope between 2005 and 2012. Some of these data were presented by \citet{buc06} and \citet{cor11}. Maps of the integrated HC$_3$N $J=10-9$ emission intensity towards six of our seven chosen sources (L1172, L1251A, L1389, L1495B, L1512 and TMC-1C) are shown in Figure \ref{fig:maps}. These observations were obtained using the OSO with a beam size of $42''$ and beam efficiency of $0.5\pm0.05$. A $30''$ map spacing was used, which was resampled to a $7.5''$ per pixel grid using bilinear interpolation.  For TMC-1, HC$_3$N maps were published by \citet{hir92} and \citet{pra97}.

To search for C$_6$H and C$_6$H$^-$, we targeted the strongest HC$_3$N peak positions in our maps of L1389, L1495B, L1512 and TMC-1C.  For L1172 we targeted the sub-mm core cataloged by \citet{dif08}, which fell within 15$''$ of the HC$_3$N peak. Our chosen L1251A position does not coincide with the main HC$_3$N peak because it was based on an earlier, lower-sensitivity map than that shown in Figure \ref{fig:maps}. For TMC-1, we targeted the cyanopolyyne peak (CP) where \citet{agu08} previously attempted to detect C$_4$H$^-$. The adopted coordinates for our anion survey targets are listed in Table \ref{tab:sources}.

Our L1172 and L1512 C$_6$H$^-$ positions overlap the prestellar cores identified by \citet{vis02} and \citet{war94}, respectively. For L1389, the HC$_3$N emission maps out a very compact clump, coincident with the center of the Bok Globule CB17 \citep{cle88}. The HC$_3$N peak also matches closely the location of strongest microwave/sub-mm continuum emission observed by \citet{lau10} and denoted L1389 SMM1 (the position of which is indicated in the bottom-left panel of Figure \ref{fig:maps} by the cross inside the dashed black circle of the GBT beam). As shown in Figure \ref{fig:maps}, our targeted L1251A position is $\approx40''$ south-east of the center of the protostar L1251 IRS3 \citep{lee10}, although the outer envelope of this protostar probably intersects the GBT beam to some extent, as discussed by \citet{cor11}.

From dust continuum emission, \citet{lau10} obtained a number density for L1389 SMM1 of $6\times10^6$~cm$^{-3}$. \citet{pav06} derived a moderately high CO depletion factor ($\sim40$), and inferred a relatively evolved chemical age of $\sim2$~Myr for this collapsing core. The core was recently found to contain a very young, low-luminosity protostar (referred to as CB17~MMS; \citealt{che12}), $8''$ NE of our targeted GBT beam center, as well a more evolved Class 0/I protostar $20''$ NW. Given that the radii of typical low-mass protostellar envelopes are $\sim10^4$~AU (\emph{e.g.} \citealt{jor02}), at a distance of 250~pc, our targeted L1389 beam is likely to be dominated by protostellar envelope matter.

L1495B, on the other hand, is a quiescent molecular cloud with no known protostars nearby. It appears to be chemically young, with an apparently low level of CO depletion consistent with less-evolved interstellar clouds \citep{hir04}. In TMC-1C, \citet{buc06} identified an anti-correlation between the spatial distributions of HC$_3$N and C$^{18}$O, at least partly attributable to CO depletion. Thus, our observations of TMC-1C (at the HC$_3$N peak) probably sample more chemically-evolved, depleted gas than L1495B. No protostars are known to be present in our observed TMC-1C beam, but several sub-mm sources are located in the surrounding cloud (indicated by white crosses in the lower-right panel of Figure \ref{fig:maps}), showing this to be a region of active star formation. South-west of TMC-1C lies the source TMC-1 (CP), which is a well-known chemically-young dark cloud, with only a modest degree of depletion \citep{hir92,cor12}.

In Table \ref{tab:sources}, we provide a basic categorization of our target sources in light of their properties described above: `quiescent' refers to those cloud cores that are chemically young, show no evidence for active star-formation within the telescope beam (such as outflows or compact sub-mm/IR emission), and do not appear to be undergoing collapse; `star-forming' is used for clouds showing nearby active star-formation and a greater degree of chemical evolution; finally, `protostellar' describes our L1389 position, which contains the low-luminosity protostar CB17~MMS.

\section{Anion Observations}
\label{obs}

Observations of emission lines of C$_6$H ($J=10.5-9.5$), C$_6$H$^-$ ($J=10-9$ and $11-10$) and HC$_3$N ($J=3-2$ and $4-3$) were carried out between 2010 April and 2011 January using the Ka receiver of the NRAO 100-meter Robert C. Byrd Green Bank Telescope\footnote{The National Radio Astronomy Observatory is a facility of the National Science Foundation operated under cooperative agreement by Associated Universities, Inc.} (GBT). The GBT Spectrometer was used with a bandwidth of 50 MHz and $8192\times6.1$~kHz channels (corresponding to a velocity spacing of $\approx0.065$ \kms). Four spectral windows were used, which allowed simultaneous observation of the C$_6$H and C$_6$H$^-$ lines.  For the compact, spatially-isolated source L1512, beam switching (with $78''$ throw) was used, and for all other targets, frequency switching was used. Pointing was checked every one to two hours and was typically accurate to within $5''$. In the middle of the observed frequency range (28 GHz), the telescope beam FWHM was $26''$ and the main beam efficiency was 0.90. Total system temperatures were typically in the range $60-80$~K and the zenith opacity was $0.05\pm0.02$. Intensity calibration was performed using beam-switched observations of the compact radio source NGC\,7027. Measured antenna temperatures were subsequently corrected for opacity, spillover, ohmic loss, blockage efficiency and beam efficiency, then averaged using standard GBTIDL routines.

We obtained observations of C$_4$H ($N = 2-1$, $J = 1.5-0.5$, $F = 2-1$) and C$_4$H$^-$ ($J=2-1$) between 2011 December and 2012 February at around 19~GHz using the GBT K-band focal-plane array (KFPA). These observations were obtained with channel spacing 12.2~kHz, beam efficiency 0.92, beam size 39$''$ and zenith opacity $0.025\pm0.015$. 

Additional single-pointing HC$_3$N $J=10-9$ spectra of our sources were obtained using the OSO at our adopted target positions. A list of all the observed transitions is given in Table \ref{tab:trans}.

\section{Results}
\label{results}

\subsection{New Anion Detections}

\begin{deluxetable*}{llcccc}
\centering
\tabletypesize{\footnotesize}
\tablecaption{Observed species, transitions and frequencies\label{tab:trans}}
\tablewidth{0pt}
\tablehead{
Species&Transition&Frequency&Ref.&Telescope&HPBW\tablenotemark{a}\\
&&(MHz)&&&($''$)\\
}
\startdata
C$_4$H & $N = 2-1$, $J = 1.5-0.5$, $F = 2-1$ &19054.4760&1&GBT&38\\
C$_4$H$^-$ & $J=2 - 1$               &18619.7580&2,3&GBT&39\\
C$_6$H$^-$ & $J=10 - 9$              &27537.1302&4&GBT&26\\
C$_6$H$^-$ & $J=11 - 10$             &30290.8133&4&GBT&24\\
C$_6$H& $J=10.5 - 9.5, f, F=11 - 10$ &29109.6437&5&GBT&25\\
C$_6$H& $J=10.5 - 9.5, f, F=10 - 9$  &29109.6855&5&GBT&25\\
C$_6$H& $J=10.5 - 9.5, e, F=11 - 10$ &29112.7087&5&GBT&25\\
C$_6$H& $J=10.5 - 9.5, e, F=10 - 9$  &29112.7503&5&GBT&25\\
HC$_3$N& $J=3 - 2, F=3 - 3$          &27292.9018&6&GBT&27\\
HC$_3$N& $J=3 - 2, F=2 - 1$          &27294.0758&6&GBT&27\\
HC$_3$N& $J=3 - 2, F=3 - 2$          &27294.2927&6&GBT&27\\
HC$_3$N& $J=3 - 2, F=4 - 3$          &27294.3451&6&GBT&27\\
HC$_3$N& $J=3 - 2, F=2 - 2$          &27296.2334&6&GBT&27\\
HC$_3$N& $J=4 - 3, F=4 - 4$          &36390.8861&6&GBT&20\\
HC$_3$N& $J=4 - 3, F=3 - 2$          &36392.2358&6&GBT&20\\
HC$_3$N& $J=4 - 3, F=4 - 3$          &36392.3293&6&GBT&20\\
HC$_3$N& $J=4 - 3, F=5 - 4$          &36392.3630&6&GBT&20\\
HC$_3$N& $J=4 - 3, F=3 - 3$          &36394.1765&6&GBT&20\\
HC$_3$N& $J=10-9, F=9-8$             &90978.9838&6&OSO&42\\
HC$_3$N& $J=10-9, F=10-9$            &90978.9948&6&OSO&42\\
HC$_3$N& $J=10-9, F=11-10$           &90979.0024&6&OSO&42
\enddata
\tablenotetext{a}{Telescope half-power beam-width at observed frequency.}
\tablerefs{(1) \citet{got83}; (2) \citet{gup07}; (3) \citet{mcc08}; (4) \citet{mcc06}; (5) \citet{mcc05}; (6) HC$_3$N line predictions are based on spectroscopic data summarized by \citet{tho00} and \citet{dez71}.}
\end{deluxetable*}

\begin{figure*}
\centering
\includegraphics[width=0.88\columnwidth]{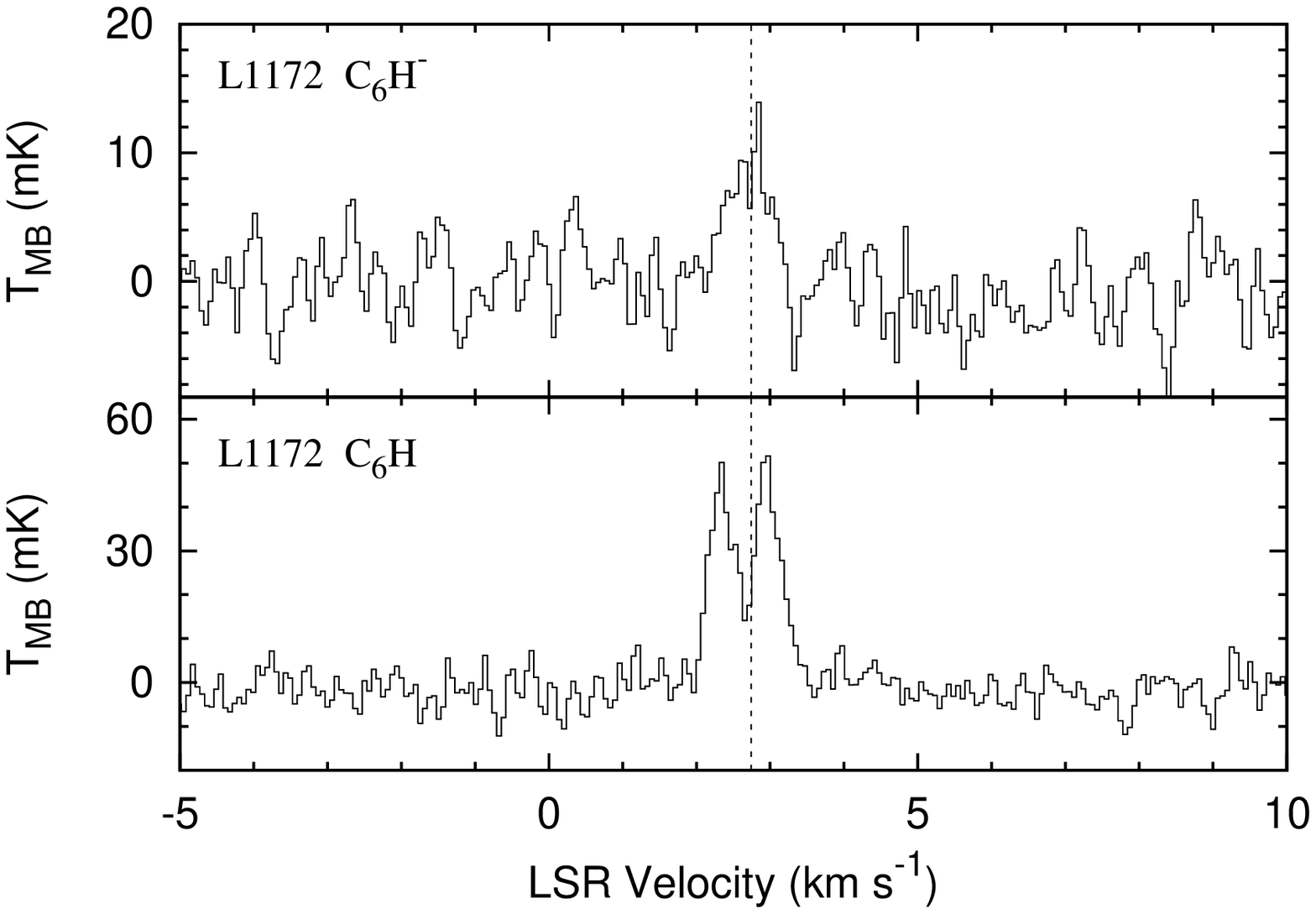}
\hspace{3mm}
\includegraphics[width=0.88\columnwidth]{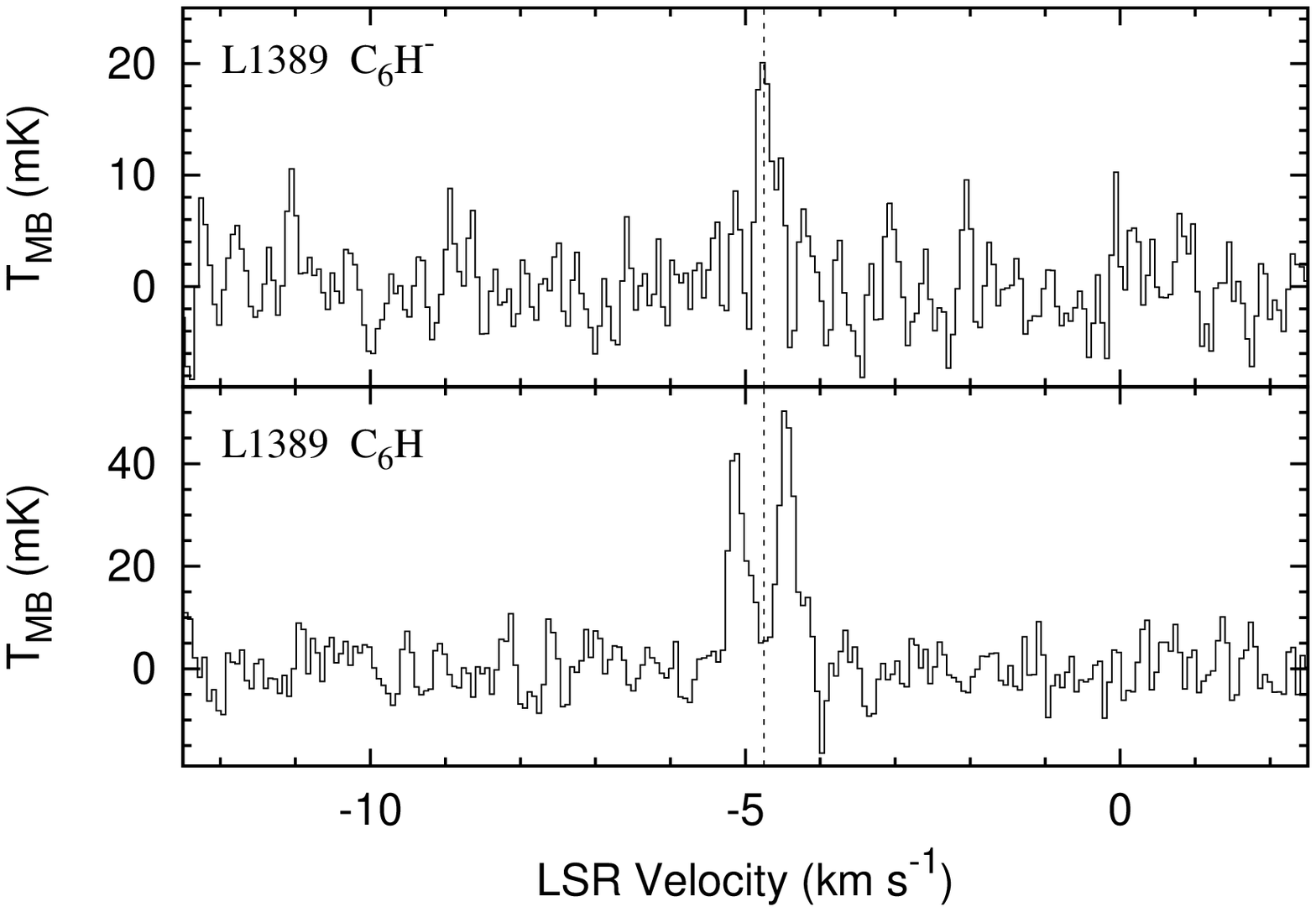}
\\
\includegraphics[width=0.88\columnwidth]{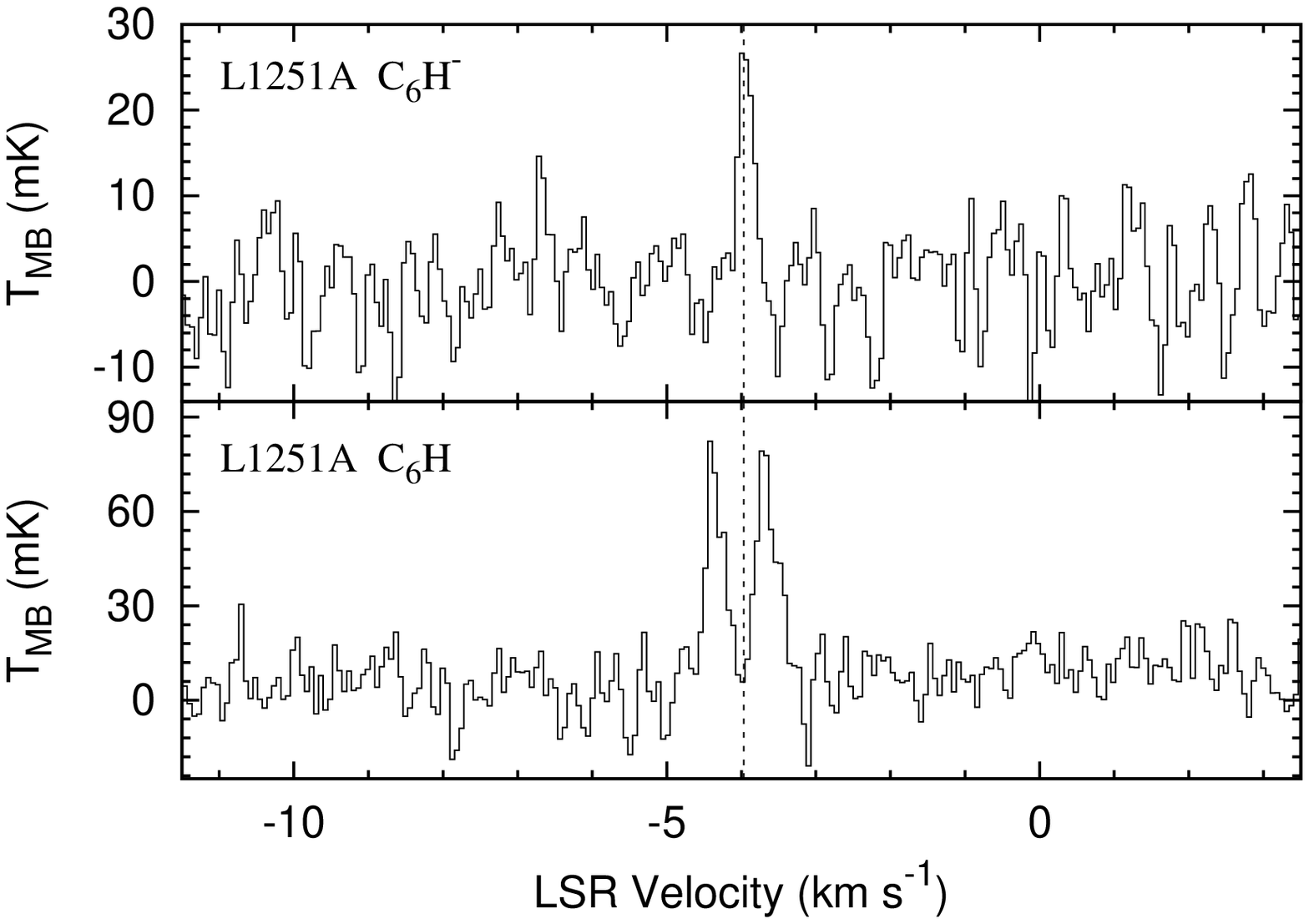}
\hspace{3mm}
\includegraphics[width=0.88\columnwidth]{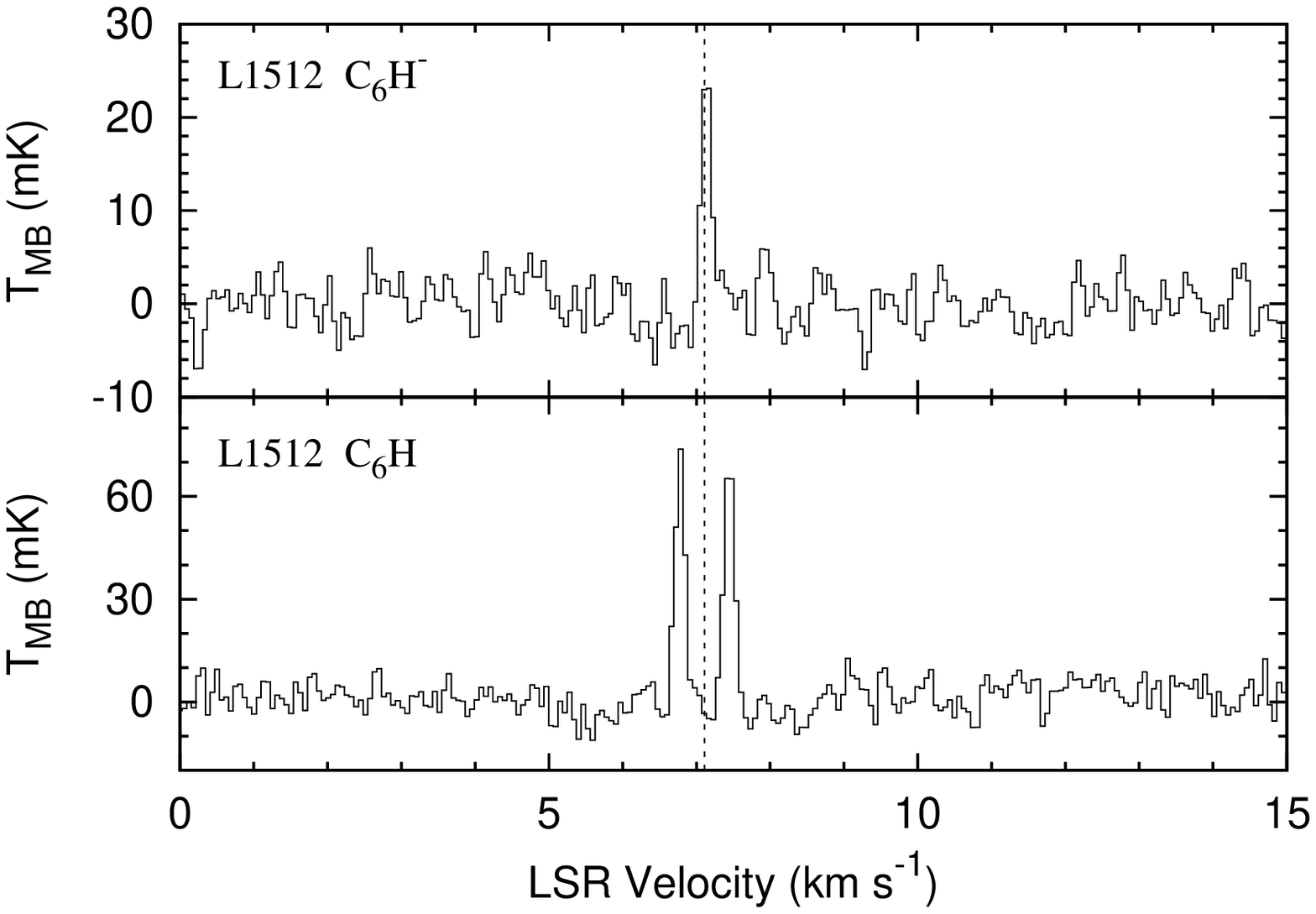}
\\
\includegraphics[width=0.88\columnwidth]{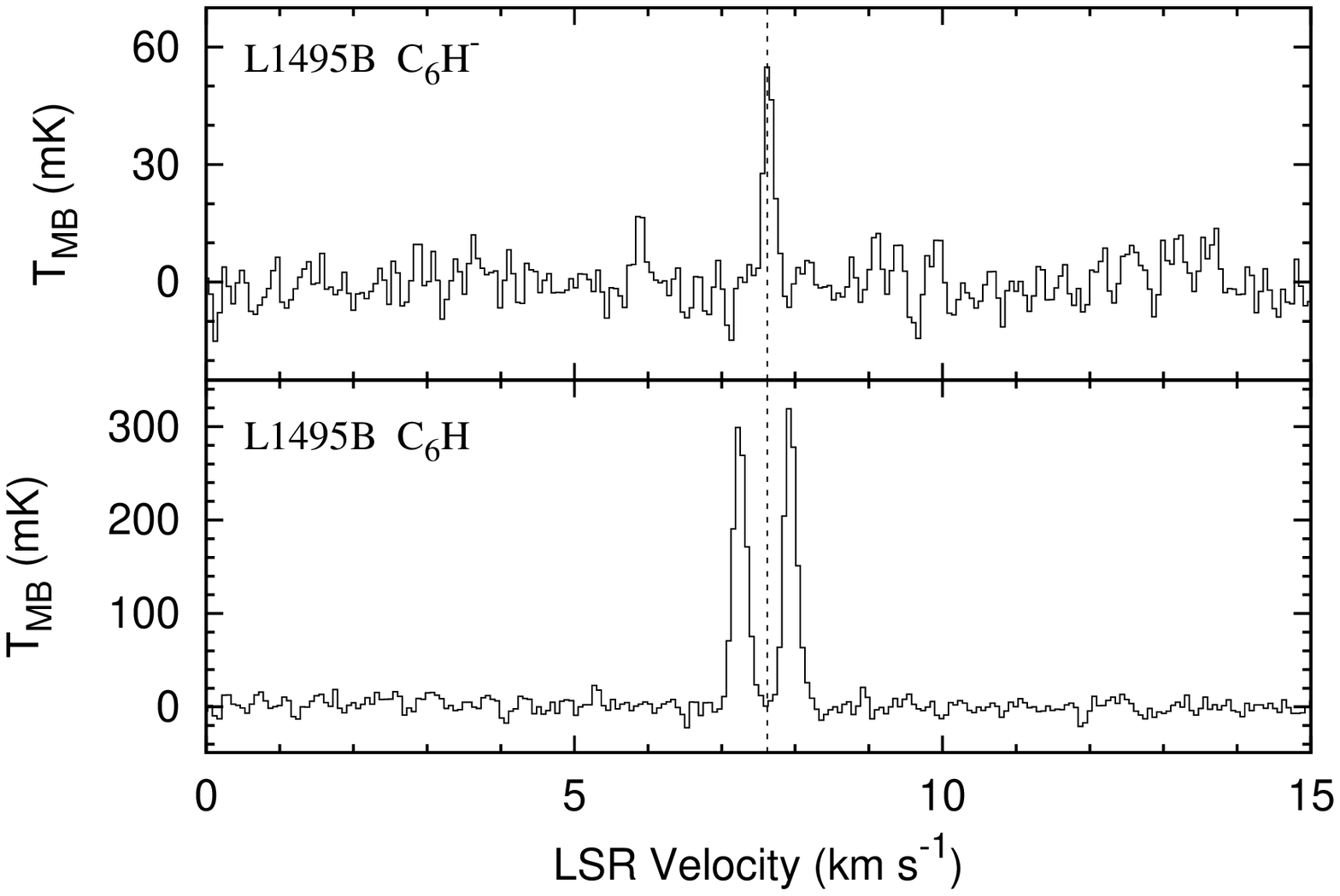}
\hspace{3mm}
\includegraphics[width=0.88\columnwidth]{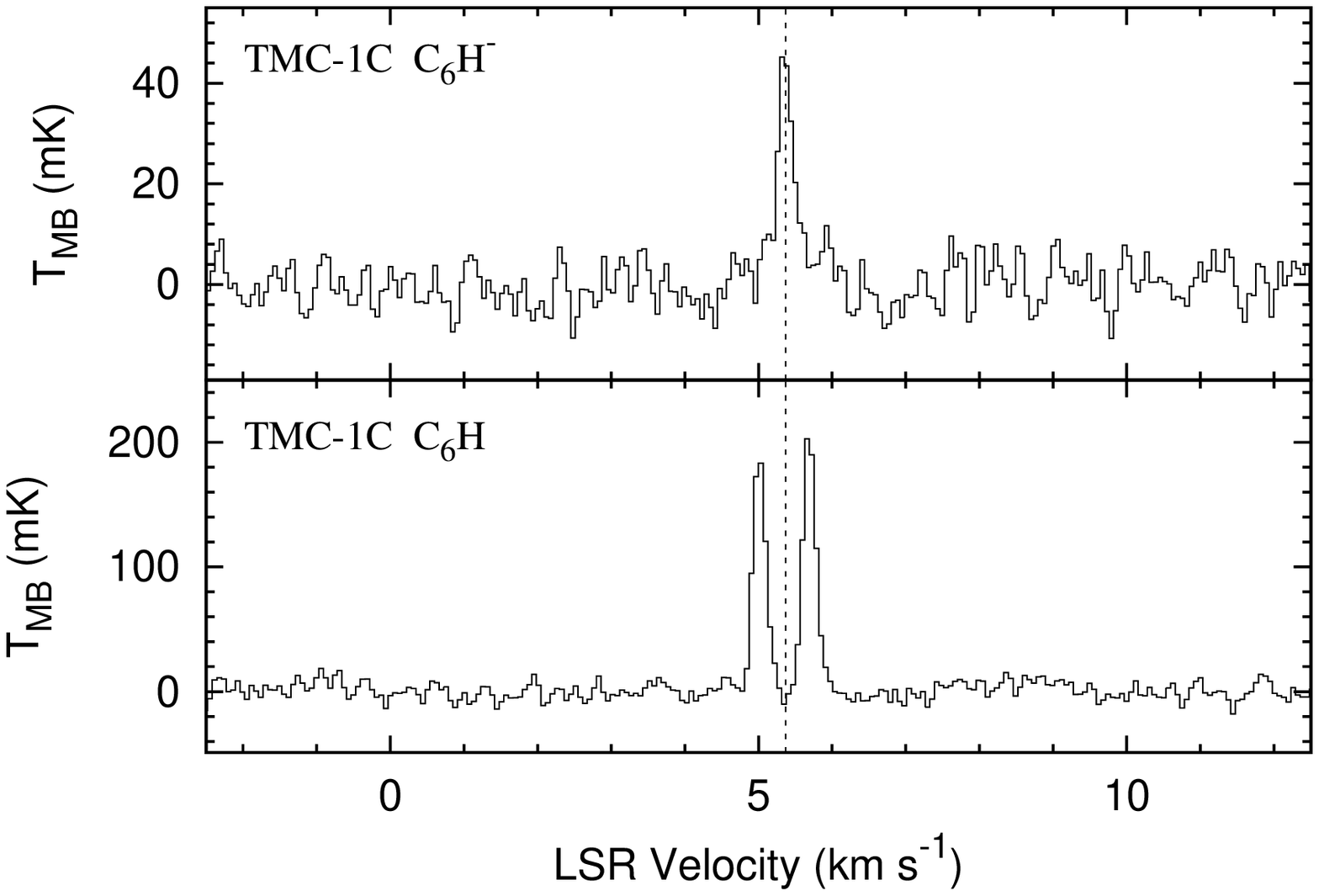}
\\
\includegraphics[width=0.88\columnwidth]{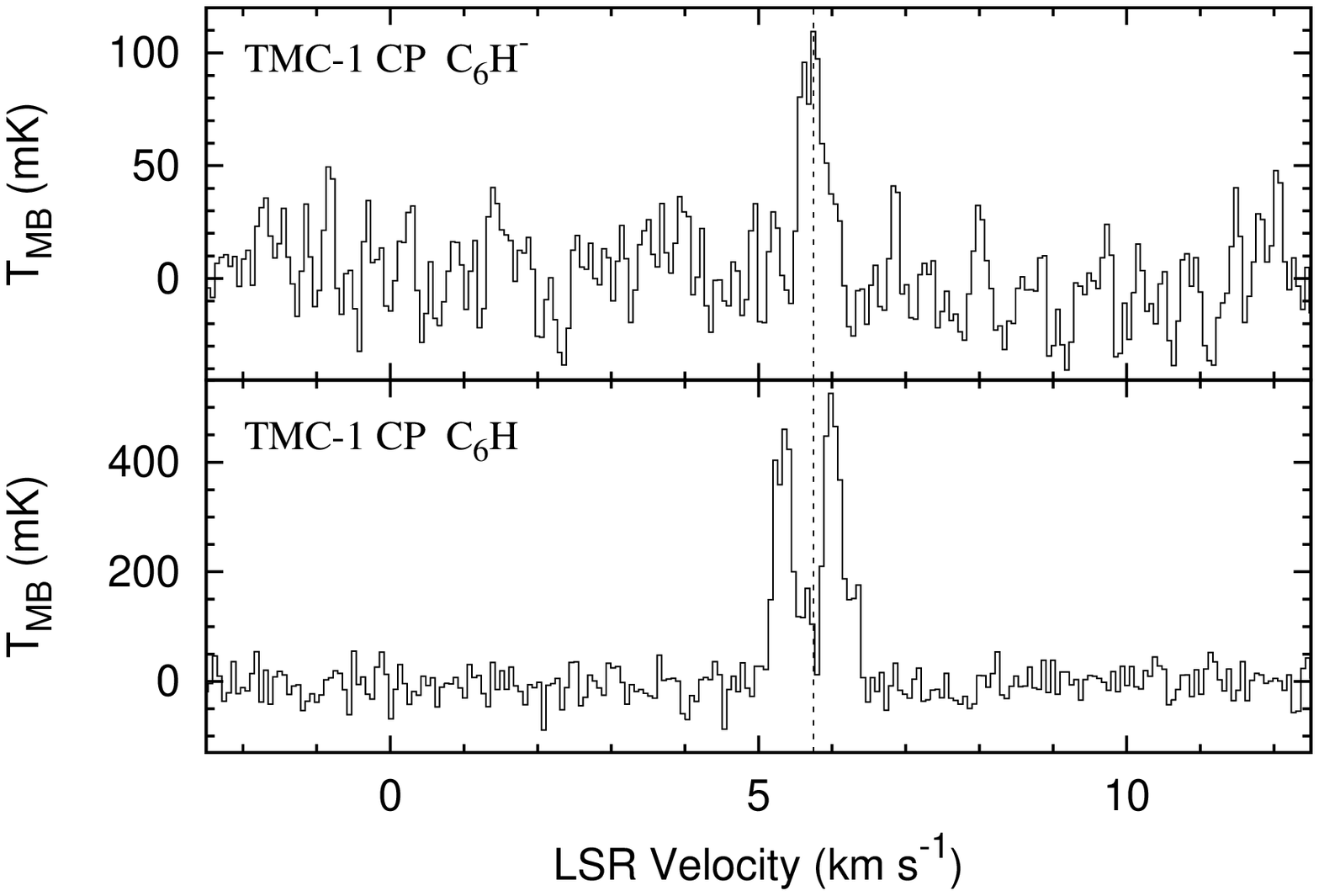}
\hspace{3mm}
\includegraphics[width=0.88\columnwidth]{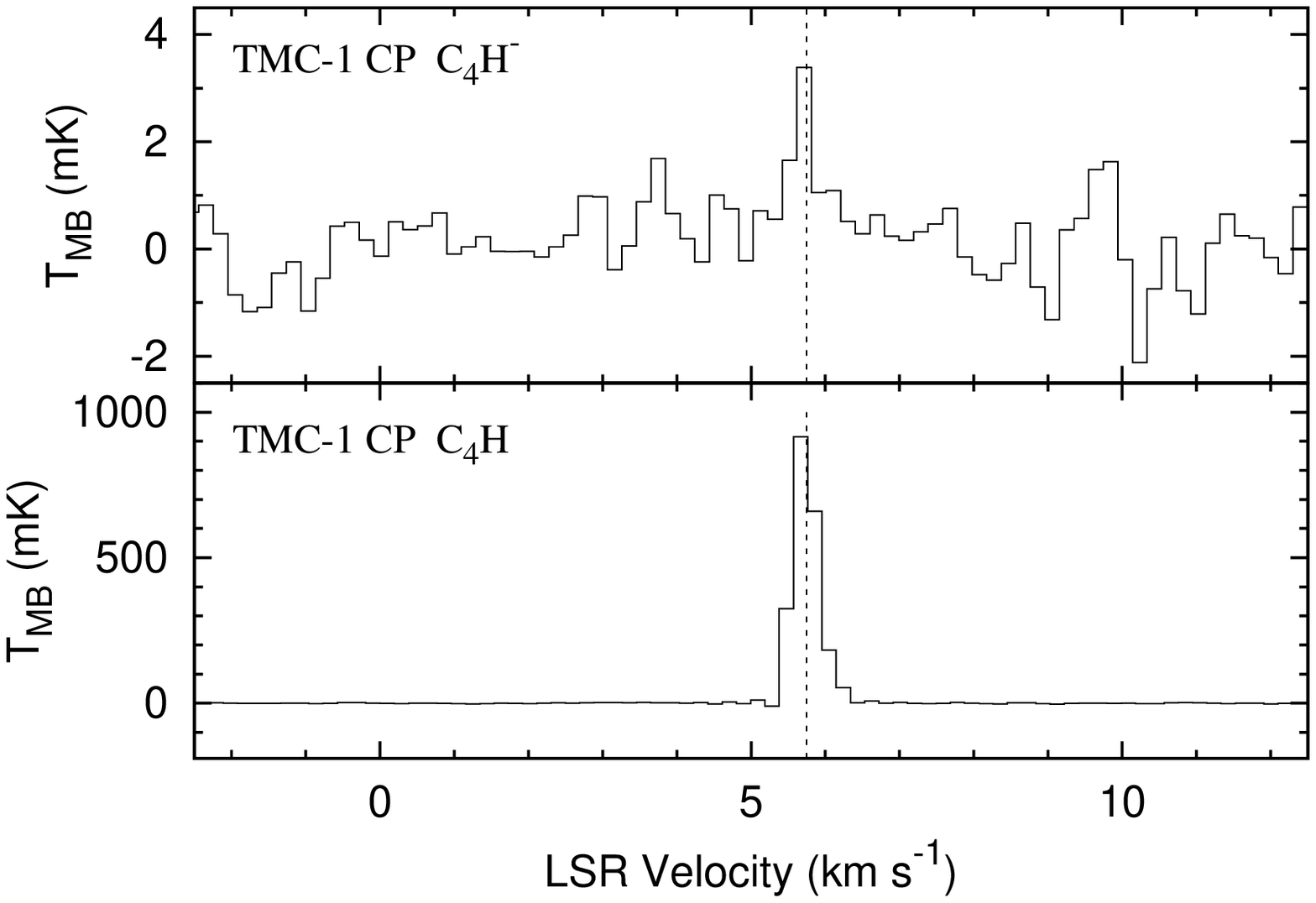}
\caption{Observed C$_6$H$^-$ spectra (average of $J=10-9$ and $J=11-10$ transitions), and C$_6$H $J=10.5-9.5,\,f$ spectra. The C$_6$H velocity scale is given with respect to the (weighted) mean frequency of the two hyperfine components. For TMC-1, observed C$_4$H$^-$ ($J=2-1$) and C$_4$H ($N = 2-1, J = 1.5-0.5, F = 2-1$) spectra are also shown (bottom-right panel). Dashed vertical lines indicate the least-squares HC$_3$N velocities for each target. \label{fig:spectra}}
\end{figure*}

\begin{deluxetable*}{lccccc}
\centering
\tabletypesize{\footnotesize}
\tablecaption{Molecular line measurements and anion-to-neutral ratios\label{tab:results}}
\tablewidth{0pt}
\tablehead{
Source&$\int{T_{MB}dv}$(C$_6$H$^-$)\tablenotemark{a}&$\int{T_{MB}dv}$(C$_6$H)\tablenotemark{b}&$N$(C$_6$H$^-$)&$N$(C$_6$H)&C$_6$H$^-$/C$_6$H\\
& (mK\,km\,s$^{-1}$)&(mK\,km\,s$^{-1}$)&($10^{10}$ cm$^{-2}$)&($10^{11}$ cm$^{-2}$)&(\%)
}
\startdata
L1172& \phantom{1}6.7 (0.8) &\phantom{1}41.1 (1.6) &\phantom{1}2.4 (0.3) &\phantom{1}7.1 (0.3) &3.3 (0.5)\\
L1251A    & \phantom{1}6.5 (1.0) &\phantom{1}43.6 (2.1) &\phantom{1}2.3 (0.4) &\phantom{1}7.6 (0.4) &3.0 (0.6)\\
L1389&\phantom{1}5.9 (0.8) &\phantom{1}27.1 (1.3) &\phantom{1}2.1 (0.3) &\phantom{1}4.7 (0.2) &4.4 (0.8)\\
L1495B    &\phantom{1}9.6 (1.0) &141.6 (2.2)&\phantom{1}3.4 (0.4) &24.6 (0.4)&1.4 (0.2)\\
L1512     &\phantom{1}4.3 (0.4) &\phantom{1}26.3 (0.9) &\phantom{1}1.5 (0.1) &\phantom{1}4.6 (0.2) &3.3 (0.4)\\
TMC-1C    &13.6 (1.1)&\phantom{1}88.1 (1.7) &\phantom{1}4.8 (0.4) &15.3 (0.3)&3.1 (0.3)\\
TMC-1 CP  &41.6 (5.0)&332.4 (9.0)&14.7 (1.8)&57.8 (1.6)&2.5 (0.4)
\enddata
\tablenotetext{a}{Integrated intensities of average C$_6$H$^-$ $J=10-9$ and $J=11-10$ spectra.}
\tablenotetext{b}{Integrated intensities summed over both hyperfine components of the C$_6$H $J=10.5-9.5,\,f$ transition.}
\tablecomments{Uncertainties given in parentheses are $\pm68$\% Monte Carlo errors. Measurements for L1251A and L1512 supersede those of \citet{cor11} as explained in Section \ref{results}}.
\end{deluxetable*}

The C$_6$H$^-$ anion and its neutral counterpart C$_6$H were detected in all of the surveyed sources. This is the first time anions have been detected in L1172, L1389, L1495B and TMC-1C. The observed C$_6$H$^-$ and C$_6$H spectra are shown in Figure \ref{fig:spectra}.

Spectra of the $J=10-9$ and $J=11-10$ transitions of C$_6$H$^-$  were averaged in velocity space to improve the signal-to-noise ratio; integrated intensities of the averaged spectra are given in Table \ref{tab:results}.  Due to partial blending of the hyperfine components of the C$_6$H ($J=10.5-9.5,\,f$) lines, the integrated intensity summed over both components is given in Table \ref{tab:results}. Given the weakness of these lines, they can safely be assumed to be optically thin, and Equation 2 of \citet{lis02} was used to derive column densities for C$_6$H and C$_6$H$^-$ (given in Table \ref{tab:results}). Spectroscopic data for the transitions of interest (Table \ref{tab:trans}) were obtained from the Cologne Database for Molecular Astronomy \citep{mul05}, and an excitation temperature of 10~K was assumed. Error estimates were derived using 500 Monte Carlo noise replications for each measurement.

In TMC-1, C$_6$H excitation temperatures of 5.2~K and 6.7~K were measured by \citet{bel99} and \citet{sak07}, respectively. Such sub-thermal excitation might be expected given the large Einstein $A$ coefficients for rotational transitions of this molecule, which are a direct consequence of its large (5.5~D) dipole moment. However, the value obtained by \citet{bel99} may be subject to uncertainty due to the effects of telescope beam dilution, and the \citet{sak07} value may not be applicable to our observations because it was obtained using the $J=16.5-15.5$ transition, which has an Einstein $A$ coefficient about four times greater than that of the $J=10.5-9.5$ transition we observed. The larger Einstein $A$ results in an increased likelihood of sub-thermal excitation of the $J=16.5$ level. If the C$_6$H excitation temperature is as low as 5~K, then the calculated column densities for C$_6$H and C$_6$H$^-$ will be about 36\% smaller than the values in Table \ref{tab:results}.  Due to the similar moments of inertia of C$_6$H and C$_6$H$^-$, their rotational levels occur at similar energies, so that the calculated anion-to-neutral column density ratios (C$_6$H$^-$/C$_6$H) are insensitive to small uncertainties in excitation, provided both molecules share a common excitation temperature. The mean C$_6$H$^-$/C$_6$H ratio for the seven sources is 3.0\%, with a standard deviation of 0.92\%.

Our new measurements for C$_6$H and C$_6$H$^-$ in L1251A and L1512 supersede those presented by \citet{cor11}. Although the column densities and anion-to-neutral ratios measured in the present study match those of \citet{cor11} within the stated errors, the newer values benefit from improved calibration and analysis methods and have smaller uncertainties. The difference between our TMC-1 (CP) C$_6$H$^-$/C$_6$H value of $2.5\pm0.4$\% and the value of $1.6\pm0.3$\% obtained by \citet{bru07} is just beyond the range of the one-sigma error bars, and could be the result of statistical noise in the spectra. Alternatively, because \citet{bru07} derived their value from a different set of transitions, non-LTE excitation effects in C$_6$H and/or C$_6$H$^-$ could be partially responsible for the discrepancy.

The observed C$_4$H$^-$ and C$_4$H spectra of TMC-1 (CP) are shown in the bottom-right panel of Figure \ref{fig:spectra}. The $J = 2-1$ line of C$_4$H$^-$ is detected in a single channel with a peak antenna temperature of 3.4~mK, which corresponds to 4.8$\sigma$ (where $\sigma$ is the RMS noise of the baseline-subtracted spectrum). Using a Gaussian fit to this C$_4$H$^-$ line with 1000 Monte Carlo noise replications, the central velocity was found to be $5.70\pm0.05$~km\,s$^{-1}$ with a FWHM of $0.43\pm0.13$~km\,s$^{-1}$.  Within the errors, these parameters match those derived for HC$_3$N (given in Table \ref{tab:hc3n}), which adds confidence to our C$_4$H$^-$ detection.  The integrated line intensity is $1.0\pm0.3$~mK\,km\,s$^{-1}$, which corresponds to a column density of $N({\rm C_4H}^-)=(8.0\pm2.4)\times10^9$~cm$^{-2}$ (assuming an excitation temperature of 10~K). The neutral C$_4$H line has an integrated intensity of $411.3\pm0.7$~mK\,km\,s$^{-1}$, which requires $N({\rm C_4H})=(6.91\pm0.01)\times10^{14}$~cm$^{-2}$. The corresponding anion-to-neutral ratio C$_4$H$^-$/C$_4$H is $(1.2\pm0.4)\times10^{-5}$. This value is consistent with the upper limit of $5.2\times10^{-5}$ measured by \citet{agu08}, and our C$_4$H column density also closely matches their value of $7.1\times10^{14}$~cm$^{-2}$. Both $N({\rm C_4H})$ and $N({\rm C_4H}^-)$ are relatively insensitive to the adopted excitation temperature, and vary by only 12\% over the range $5-10$~K.

\subsection{HC$_3$N Observations and Radiative Transfer Modeling}

\begin{figure*}
\includegraphics[width=2.0\columnwidth]{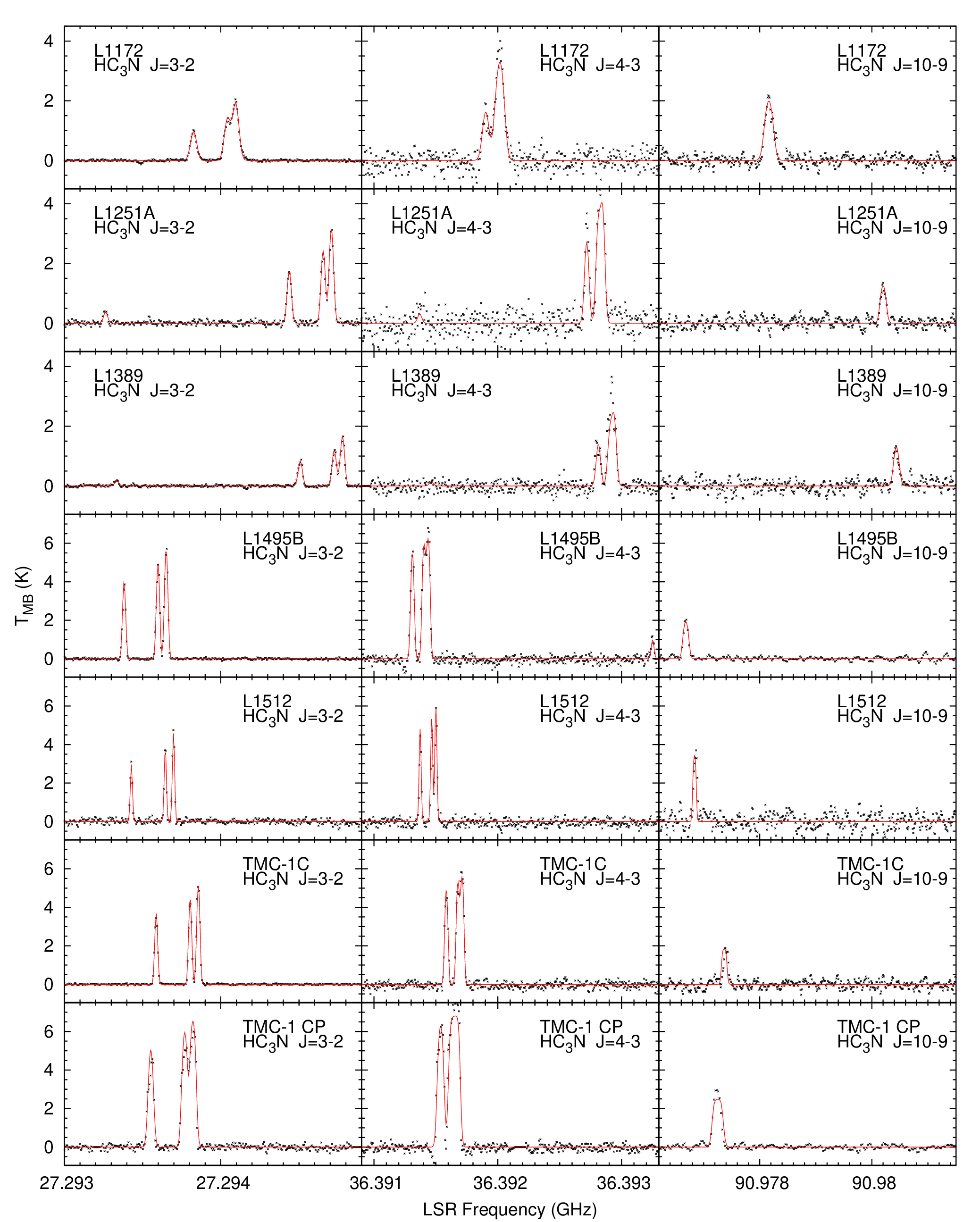}
\caption{Observed HC$_3$N spectra (black points) with RADEX least-squares models overlaid (red curves). Frequency-switching residuals have been masked from the observed data.  Three main hyperfine peaks are visible in the $J=3-2$ and $J=4-3$ spectra. Model parameters are given in Table \ref{tab:hc3n}. \label{fig:hc3n}}
\end{figure*}

\begin{deluxetable}{lcccc}
\centering
\tabletypesize{\footnotesize}
\tablecaption{HC$_3$N RADEX fit results\label{tab:hc3n}}
\tablewidth{0pt}
\tablehead{
Source&$N$(HC$_3$N)&$v$(HC$_3$N)&$\Delta{v}$(HC$_3$N)&$n_{\rm H_2}$\\
&($10^{13}$ cm$^{-2}$)&(km\,s$^{-1}$)&(km\,s$^{-1}$)&($10^{4}$ cm$^{-3}$)
}
\startdata
L1172   &\phantom{1}2.7 &\phantom{-}2.74 &0.49&7.5 (2.3)\\
L1251A      &\phantom{1}3.3 &-3.97&0.31&2.1 (0.4)\\           
L1389   &\phantom{1}1.5 &-4.75&0.32&5.2 (1.3)\\           
L1495B      &\phantom{1}8.2 &\phantom{-}7.62 &0.19&1.1 (0.3)\\
L1512       &\phantom{1}4.2 &\phantom{-}7.11 &0.10&2.6 (0.9)\\
TMC-1C      &\phantom{1}6.4 &\phantom{-}5.37 &0.18&1.1 (0.4)\\
TMC-1 CP    &19.5&\phantom{-}5.75 &0.32&1.0 (0.2)                  
\enddata
\tablecomments{Uncertainties on $n_{\rm H_2}$ (in parentheses) were calculated assuming $\pm1$~K uncertainty on the kinetic temperature, but may be larger due to the possible effects of beam dilution.}
\end{deluxetable}

Our HC$_3$N maps (Figure \ref{fig:maps}) show the presence of one or more compact molecular condensations in each source, and highlight a clumpy structure in this carbon-chain-rich gas.  The peaks in HC$_3$N emission tend to coincide approximately with the locations of known prestellar cores and protostars, the positions of which are denoted with crosses and asterisks, respectively.  As a result of the collapse and infall processes occurring in these objects, their densities are expected to be greater than the surrounding gas. This should result in increased collisional excitation of the $J=10$ level of HC$_3$N, which has a critical density of $7\times10^5$~cm$^{-3}$ \citep{buc06}. Thus, the intensity variations shown in our HC$_3$N OSO maps represent a combination of variations in gas density and total HC$_3$N column.

Spectra of three rotational transitions of HC$_3$N for each source, observed at the locations targeted by our GBT anion survey, are shown in Figure \ref{fig:hc3n}. These span a range of upper-state energy levels $1.82-16.69$~cm$^{-1}$ ($2.6-24.0$~K). In order to derive gas densities and velocities, the HC$_3$N spectra were subject to fitting using the RADEX radiative transfer code \citep{van07}. Collisional transition rates were taken from the Leiden Atomic and Molecular Database\footnote{http://www.strw.leidenuniv.nl/$\sim$moldata} \citep{sch05}, which tabulates scaled versions of the original data of \citet{gre78}. Hyperfine structure (HFS) was accounted for by assuming LTE population of the hyperfine levels in each $J$ state.  The primary collision partner number density ($n_{\rm H_2}$), molecular column density ($N$), Doppler line FWHM ($\Delta{v}$) and Doppler velocity ($v$) were optimized for the spectra of each source using the MPFIT least-squares algorithm \citep{mar08}. The best-fitting parameters are given in Table \ref{tab:hc3n}. The kinetic temperature was fixed at 10~K, then varied by $\pm1$~K to produce the quoted error estimates on $n_{\rm H_2}$. For all sources, a temperature of 10~K produced a good fit to all five HFS components (the two weakest HFS components are not shown in Figure \ref{fig:hc3n}). The HC$_3$N Doppler velocities are shown with vertical dashed lines in Figure \ref{fig:spectra} and closely match the central velocities of the detected hydrocarbons and anions. In several cases, the HC$_3$N line widths are very narrow; for L1512, $\Delta{v}=0.1$~km\,s$^{-1}$, which is only slightly greater than the 10~K thermal line-width of 0.07~km\,s$^{-1}$, indicating an unusual lack of turbulence, flows or shears in the central $\sim30''$ of this source.

The C$_6$H$^-$/C$_6$H ratio is plotted as a function of $n_{\rm H_2}$ in Figure~\ref{fig:ANR}, and shows a positive correlation. There is considerable scatter and uncertainty on $n_{\rm H_2}$, but consistent with previous studies (\citealt{hir92}, \citealt{vis02}, \citealt{lau10} and \citealt{kir05}, respectively), we find the lowest density in TMC-1, highest densities in in L1172 and L1389, and an intermediate density in L1512. Our calculated $n_{\rm H_2}$ values for L1172, L1389 and L1512 are, however, systematically lower than previously-derived values by up to two orders of magnitude. This discrepancy is likely indicative of a problem in our method, which can be attributed to the effects of different degrees of beam dilution affecting the different HC$_3$N line frequencies we observed.  Given the larger $42''$ beam size for the HC$_3$N $J=10-9$ line observed with the OSO 20-m compared to the $27''$ GBT beam for the $J=3-2$ line, it is theoretically possible that the OSO data suffers up to an additional factor of 2.6 in beam dilution compared with that of the GBT, which could result in sufficient reduction of the derived $n_{\rm H_2}$ values to account for the observed discrepancies. Figure \ref{fig:maps} (and Figure 7 of \citealt{pra97}) shows that the HC$_3$N emission from all our observed sources is distributed over an area larger than the OSO beam, which suggests that beam dilution may not be so severe. However, given the lack of spatial information on the extent of the $J=10-9$ emission on size scales less than the OSO beam size, it is impossible to quantify the importance of beam dilution in our RADEX calculations, so the $n_{\rm H_2}$ values given in Table~\ref{tab:hc3n} should be treated with caution.  More accurate estimates for $n_{\rm H_2}$ could be obtained from HC$_3$N $J=10-9$ observations using a telescope with a smaller beam (such as the IRAM 30-m), which would be a better match to that of our GBT observations.\\

\section{Discussion}

\subsection{C$_6$H$^-$}

\begin{figure}
\centering
\includegraphics[width=0.75\columnwidth,angle=270]{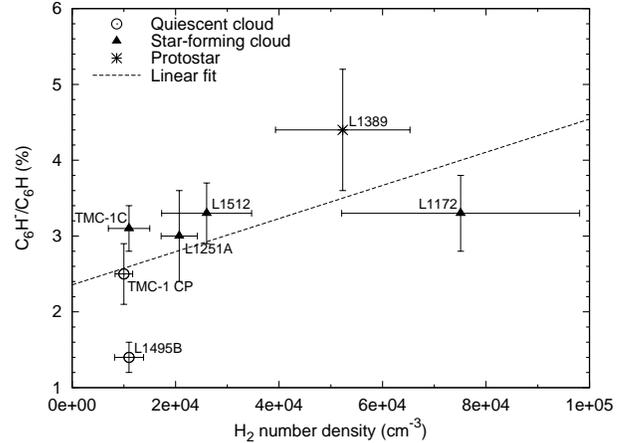}
\caption{C$_6$H$^-$ anion-to-neutral ratio \emph{vs.} H$_2$ number density ($n_{\rm H_2}$). Dashed line shows linear least-squares fit. \label{fig:ANR}}
\end{figure}

Since the 2006 discovery of C$_6$H$^-$ in TMC-1, there have been seven reported detections of this anion in various parts of the Taurus-Auriga Molecular Cloud complex (including our latest detections in L1495B, L1512 and TMC-1C). Thus far, there have been only two detections of interstellar anions outside of this small region \citep{sak10,cor11}, which highlights the importance of our new discoveries of C$_6$H$^-$ in the vicinity of L1172~SMM (a prestellar core in Cepheus) and L1389~SMM1 (a prestellar/protostellar core inside Bok Globule CB17 in Camelopardalis). These new detections confirm that anions are indeed widespread in star-forming regions outside of Taurus, and that C$_6$H$^-$ appears to be ubiquitous in dense interstellar clouds wherever its parent neutral is present.

\citet{mye88} detected an outflow in L1172 that \citet{vis02} identified as originating from a protostar (L1172 SMM1) that lies about 30$''$ south of our targeted position. They also detected a dense core (presumed prestellar), about 50$''$ SE of our target, with $N({\rm H_2})=6\times10^{21}$~cm$^{-2}$. If this column density also applies to our observed position, the C$_6$H$^-$ abundance is $4\times10^{-12}$, which is a factor of a few less than in TMC-1, but an order of magnitude greater than in L1512, L1521F and L1544 \citep[see][and references therein]{cor12}, suggesting a relatively high absolute anion abundance in L1172.

By utilizing the available physical and chemical information on the interstellar clouds for which C$_6$H$^-$ has so far been detected (as summarized for our observed targets in Section \ref{sec:targets}), it is possible to examine the C$_6$H$^-$/C$_6$H ratio with respect to each cloud's chemical and dynamical evolutionary state. The evolutionary states of the observed clouds are given in Table~\ref{tab:sources}, and are shown by the different symbols on the plot in Figure~\ref{fig:ANR}. There is a clear division in the anion-to-neutral ratio between young, `quiescent' and older, `star-forming' cores: for the quiescent, chemically-young cores L1495B and TMC-1 (CP), C$_6$H$^-$/C$_6$H~$<3$\%, and for the star-forming cores/protostars, C$_6$H$^-$/C$_6$H~$\geq3$\%. L1495B is the most chemically-young and quiescent of our sources \citep{hir04}, with no evidence for nearby star-formation, and has the lowest C$_6$H$^-$/C$_6$H ratio in our sample ($1.4\pm0.2$\%). The very low-luminosity, low-mass protostar (VeLLO) L1389~SMM1, on the other hand, evidently contains among the most evolved matter in our sample (having passed through the quiescent and star-forming stages), and has the highest anion-to-neutral ratio (C$_6$H$^-$/C$_6$H~=~$4.4\pm0.8$\%). This value is similar to the value of $4.0\pm1.0$\% found in L1521F by \citet{gup09}, which also contains a VeLLO \citep{bou06}.  Comparing our results with other sources from the literature, L1527 \citep{sak07} contains a somewhat more-evolved Class 0/I protostar, and this source has a significantly larger C$_6$H$^-$/C$_6$H ratio of $9.3\pm2.9$\%. Thus, we confirm the apparent relationship identified by \citet{cor11} between the evolutionary state of interstellar matter and its C$_6$H$^-$ anion-to-neutral ratio.  The quiescent cloud Lupus-1A observed by \citet{sak10} also fits the trend (with C$_6$H$^-$/C$_6$H~=~$2.1\pm0.6$\%), as does the very low-luminosity protostar Chameleon MMS1 (with its upper limit of C$_6$H$^-$/C$_6$H~$<10$\%; \citealt{cor12b}). The dense prestellar core L1544 may deviate, however, with a relatively low ratio of $2.5\pm0.8$\% \citep{gup09}.

The observed C$_6$H$^-$/C$_6$H trend and its correlation with $n_{\rm H_2}$ shown in Figure~\ref{fig:ANR} can be understood in the context of the theoretical study of \citet{cor12}, who identified the effects of depletion on the C$_6$H$^-$ anion-to-neutral ratio, the degree of which is related to both the density and chemical age of the cloud. Depletion occurs as atoms and molecules collide with and stick to dust grains, and thus proceeds faster in denser media. Increased depletion affects the anion-to-neutral ratio in two ways: (1) the free electron abundance goes up so that more C$_6$H$^-$ is produced by radiative electron attachment and (2) the atomic O and H abundances go down, thus reducing the anion destruction rate. Our C$_6$H$^-$ observations agree with this theory, particularly for the quiescent cores and the protostars, which are believed to lie at opposite ends of the interstellar chemical/dynamic evolutionary path. Measured anion-to-neutral ratios, however, are not sufficiently accurate to draw a clear distinction among the (moderately evolved) prestellar cores and star-forming gas clouds of L1172, L1251A, L1512 and TMC-1C. These clouds likely contain gas that spans a range of densities and degrees of chemical evolution. More accurate measurements of density, depletion and the C$_6$H$^-$/C$_6$H ratio at high spatial resolution will be required in order to confirm the utility of the anion-to-neutral ratio as a measure of the evolutionary state of interstellar clouds.

\subsection{C$_4$H$^-$}

Our detection of C$_4$H$^-$ constitutes the smallest reported column density for this molecule in any source to-date. Consistent with the trend described above for C$_6$H$^-$, the C$_4$H$^-$ anion-to-neutral ratio of $(1.2\pm0.4)\times10^{-5}$ in TMC-1 (CP) is about nine times smaller than observed in L1527 by \citet{agu08}. Theory regarding C$_4$H$^-$ chemistry is less well understood than for C$_6$H$^-$, but it seems plausible to again ascribe the lower value in TMC-1 (and the moderately low value of $(8.8\pm5.3)\times10^{-5}$ in Lupus-1A; \citealt{sak10}), to chemical effects resulting from the lower densities and younger evolutionary states of these sources compared with L1527.

The very low observed C$_4$H$^-$ anion-to-neutral ratios compared with C$_6$H$^-$ are at variance with chemical models that consider the formation of C$_4$H$^-$ to be by radiative electron attachment to C$_4$H at the rate calculated by \citet{her08}.  Recent ab-initio calculations by V. Kokoouline and colleagues (private communication, 2012) show that the C$_4$H radiative attachment rate may be several orders of magnitude less than previously thought, and based on observations of the C$_4$H$^-$/C$_4$H ratio in L1527, \citet{agu08} calculated a C$_4$H radiative attachment rate a factor of 122 less than the theoretical value of \citet{her08}. However, using the laboratory data of \citet{eic07}, \citet{cor12} identified that a significant pathway to smaller hydrocarbon anions is \emph{via} reactions of larger anions with atomic oxygen. Thus, an important source of C$_4$H$^-$ is the reaction

\begin{equation}
\label{equ:c4h-}
{\rm C_5H^-} + {\rm O} \longrightarrow {\rm C_4H^-} + {\rm CO}.
\end{equation}

As a consequence of this (and the analogous reaction that forms C$_5$H$^-$ from C$_6$H$^-$ + O), using the model of \citet{cor12} at $n_{\rm H_2}=10^4$~cm$^{-3}$ (applicable to TMC-1), it is possible to set the radiative attachment rates for both C$_4$H and C$_5$H to zero and still produce an absolute C$_4$H$^-$ abundance and C$_4$H$^-$/C$_4$H ratio that are within a factor of a few of the observed values. It is therefore plausible that radiative electron attachment is not the dominant route to the formation of C$_4$H$^-$ in molecular clouds.

\section{Conclusion}

We conducted a search for the carbon-chain anion C$_6$H$^-$ in seven nearby molecular clouds selected based on their strong HC$_3$N emission. C$_6$H$^-$ and C$_6$H were detected in all sources, with a mean anion-to-neutral ratio of 3.0\% (standard deviation 0.92\%). Combined with the four previous C$_6$H$^-$ detections in other sources, we deduce that anions are ubiquitous in the ISM wherever sufficient C$_6$H is present to be easily detectable.

From the combined sample of eleven known interstellar sources with C$_6$H$^-$ detections, we confirm the trend identified by \citet{cor11} for anion-to-neutral ratios to be smaller in younger, less dense environments and larger in older, denser environments. This is consistent with the theory of \citet{cor12} that anion-to-neutral ratios are sensitive to the depletion of atomic O and H, and to the electron density, both of which become larger in denser, more chemically-evolved cores.  Molecular hydrogen number densities were derived for each source based on radiative transfer fits to observed HC$_3$N spectra; a positive correlation is observed between C$_6$H$^-$/C$_6$H and $n_{\rm H_2}$.

We report the first detection of C$_4$H$^-$ in TMC-1, and derive an anion-to-neutral ratio C$_4$H$^-$/C$_4$H~$=(1.2\pm0.4)\times10^{-5}$ ($=0.0012\pm0.0004$\%).  This is the smallest anion-to-neutral ratio yet observed and confirms a problem with the theoretical value for the C$_4$H electron attachment rate, which is apparently several orders of magnitude too large.  The observed C$_4$H$^-$ may instead be formed by reactions of the larger hydrocarbon anions (C$_6$H$^-$ and C$_5$H$^-$) with atomic oxygen.

Given the widespread presence of anions in space, a combination of further dedicated laboratory, observational and theoretical studies will be required in order to better understand their behavior and thus their full importance in astrochemistry. Measurements of radiative electron attachment rates for hydrocarbons, and branching ratios for reactions of anions with atomic oxygen will be of particular importance in this regard.

\acknowledgments
This research was supported by the NASA Exobiology Program and the NASA Astrobiology Institute through the Goddard Center for Astrobiology. We gratefully acknowledge the assistance of Tom Millar and Catherine Walsh during the 2010 Onsala HC$_3$N observing run. JVB thanks John Richer for support.

\end{document}